\documentclass[10pt,twocolumn]{IEEEtran}
\usepackage{amsmath}
\usepackage{multirow}
\usepackage{graphicx}
\usepackage{tikz}
\usepackage{textcomp}
\usepackage{hyperref}

\newcommand\copyrighttext{%
  \footnotesize 
This is the author's version of an article that has been published in IEEE Communications Letters, 23(5), pp. 802 - 805. Changes were made to this version by the publisher prior to publication.
The final version of record is available at \href{https://doi.org/10.1109/LCOMM.2019.2906178}{https://doi.org/10.1109/LCOMM.2019.2906178}

Copyright (c) 2019 IEEE. Personal use of this material is permitted. Permission from IEEE must be obtained for all other uses by emailing 
pubs-permissions@ieee.org.}
\newcommand\copyrightnotice{%
\begin{tikzpicture}[remember picture,overlay]
\node[anchor=north,xshift=50pt,yshift=-7pt] at (current page.north) {\fbox{\parbox{\dimexpr0.7\textwidth-\fboxsep-\fboxrule\relax}{\copyrighttext}}};
\end{tikzpicture}%
}

\begin{document}

\title{Strengths and Weaknesses of the ETSI Adaptive DCC Algorithm: A Proposal for Improvement }

\author{Ignacio Soto, Oscar Amador, Manuel Urueña, and~Maria Calderon
\thanks{This work was partially supported by the Spanish Ministerio de Economía y Competitividad through the Texeo project (TEC2016-80339-R). 

I. Soto, O. Amador, and M. Calderon are with Departamento de Ingeniería Telemática, Universidad Carlos III de Madrid, Spain (e-mails: isoto@it.uc3m.es,  oamador@pa.uc3m.es, maria@it.uc3m.es). 
M. Urueña is with Universidad Internacional de la Rioja, Spain (e-mail: muruenya@gmail.com).

}}%

\markboth{IEEE Communications Letters}%
{Soto \MakeLowercase{\textit{et al.}}: Strengths and Weaknesses of the Adaptive DCC ETSI Standard: A Proposal for Improvement}
%
\maketitle
\copyrightnotice
%
\begin{abstract}
This letter studies the adaptive Decentralized Congestion Control (DCC) algorithm defined in the ETSI TS 102 687 V1.2.1 specification. We provide insights on the parameters used in the algorithm and explore the impact of those parameters on its performance. We show how the algorithm achieves good average medium utilization while protecting against congestion, but we also show how the chosen parameters can result in slow speed of convergence and long periods of unfairness in transitory situations. Finally, we propose a modification to the algorithm which results in significant improvements in speed of convergence and fairness.     
\end{abstract}


\IEEEpeerreviewmaketitle

\section{Introduction}

\IEEEPARstart{T}{he} European Telecommunications Standards Institute (ETSI) has developed specifications, known as \mbox{ITS-G5}~\cite{etsien302663_v1.2.1}, to support vehicle communications in the 5.9 GHz frequency band. \mbox{ITS-G5} reuses other communications standards and, in particular, it is based on the mode of operation of IEEE 802.11 in which stations work outside the context of a basic service set (originally known as 802.11p). 

ITS-G5 includes a Decentralized Congestion Control (DCC) mechanism to avoid congestion in the used wireless channel. DCC algorithms try to control the channel occupancy with the cooperative action of the different Intelligent Transport System Stations (ITS-Ss), such as vehicles or road-side units, transmitting in a channel. The aim is to ensure that the shared medium is in a state in which communications are effective. Each ITS-S can modify different parameters, or combination of parameters, to limit the channel occupancy. Examples of these parameters are the transmit power \cite{torrent2009} and the employed medium data rate, but the one that has received more attention in ETSI is the transmission rate, i.e., the time between the transmissions of consecutive messages.

The original ETSI DCC specification included a reactive algorithm, based on a state machine. In this DCC reactive algorithm~\cite{autolitano2013}, the current state is decided according to the medium occupancy, and each state defines the parameters that must be used to send data and, in particular, the maximum transmission rate. Several works in the literature have studied the performance of the DCC reactive algorithm and proposed alternatives. A key proposal is LIMERIC~\cite{bansal2013}, an adaptive DCC algorithm that has inspired many later works. Several studies have shown~\cite{bansal2014,rostami2016} that the LIMERIC algorithm performs better than the ETSI reactive DCC algorithm, or have shown limitations in the reactive algorithm~\cite{kuk2014}. In fact, the last version of the ETSI DCC specification~\cite{etsits102.687_v1.2.1} keeps the reactive algorithm but adds an adaptive algorithm based on LIMERIC.  

The behavior of the LIMERIC algorithm and, therefore, of the ETSI adaptive DCC algorithm, is regulated by several parameters. Different values of these parameters have been explored to try to achieve the best possible performance. In this letter, we provide insights on the chosen values for the parameters of the ETSI adaptive DCC algorithm, the strengths of the resulting behavior, but also its weaknesses. Then, we propose a modification to the algorithm that results in a significant improvement in speed of convergence and fairness in transitory situations.

\section{ETSI Adaptive DCC Algorithm}

\subsection{Description of the ETSI Adaptive DCC Algorithm}

The ETSI adaptive DCC algorithm~\cite{etsits102.687_v1.2.1} is used by each \mbox{ITS-S} to determine its allowed sending rate, which is represented by the parameter $\delta$: the maximum fraction of time the \mbox{ITS-S} can transmit on the channel. To calculate $\delta$, each ITS-S measures every 100ms the Channel Busy Ratio ($CBR$): the percentage of time the channel is busy in those 100ms. The difference between the measured CBR and a target CBR is the feedback to determine the proper $\delta$ in an ITS-S. The algorithm uses a smoothed version of the $CBR$, calculated every 200ms, following equation~\ref{eq:cbr}:

\begin{equation}
\label{eq:cbr}
CBR_{s}(n)=0.5 \times CBR_{s}(n-1) + 0.5 \times \frac{(CBR_{m} + CBR_{m\_p})}{2}
\end{equation}

where $CBR_{s}(n)$ is the smoothed CBR for step \textit{n}, $CBR_{s}(n-1)$ is the previous $CBR_{s}$, and $CBR_{m}$ and $CBR_{m\_p}$ are the last two measurements of the CBR (i.e., the new measurements after the previous calculation of $CBR_{s}$). Then, the ITS-S uses equation~\ref{eq:offset}, to calculate $\delta_{offset}$:

\begin{equation}
\label{eq:offset}
\delta_{offset}(n) = 
\begin{cases}
\begin{aligned}
min\{\beta \times (CBR_{t} - CBR_{s}(n)),  &G^{+}_{max}\}  \\
 \text{if } CBR_{t}&>CBR_{s}(n)  
\end{aligned} \\ 
\begin{aligned}
max\{\beta \times (CBR_{t} - CBR_{s}(n)),  &G^{-}_{min}\}  \\
 \text{if } CBR_{t} &\leq CBR_{s}(n)  
\end{aligned}
\end{cases}
\end{equation}

where $CBR_{t}$ is the target $CBR$, and $\beta$, $G^{+}_{max}$, and $G^{-}_{min}$ are parameters of the algorithm. $G^{+}_{max}$, and $G^{-}_{min}$ are meant to limit the maximum variability of $\delta_{offset}$ per step of the algorithm (i.e., to improve stability, even if the algorithm does not converge, the maximum oscillation is capped). $\delta_{offset}$ represents the needed modification in $\delta$ to keep the $CBR$ at the $CBR_{t}$ value.  $\delta_{offset}$ can be positive if $CBR_s$ is below the target, or negative otherwise. Finally, the ITS-S calculates, every 200ms, the $\delta$ at step \textit{n}, $\delta(n)$, using equations~\ref{eq:delta_wb} and \ref{eq:delta_n}:
\begin{equation}
\label{eq:delta_wb}
\delta_{wb}(n) = (1-\alpha) \times \delta(n-1) + \delta_{offset}(n)
\end{equation}
\begin{equation}
\label{eq:delta_n}
\delta(n)=
\begin{cases}
\delta_{wb}(n) & \text{if } \delta _{wb}(n)< \delta_{max} \text{ and } \delta_{wb}(n) > \delta_{min} \\
\delta_{max} & \text{if } \delta_{wb}(n) \geq \delta_{max} \\
\delta_{min} & \text{if } \delta_{wb}(n) \leq \delta_{min}
\end{cases}
\end{equation}

where $\delta(n-1)$ is the previous $\delta$, and $\alpha$ is a parameter of the algorithm. $\delta_{wb}(n)$ is a $\delta$ calculated without bounds but, as indicated by equation~\ref{eq:delta_n}, the ETSI DCC specification limits the maximum and minimum values of $\delta$ for any ITS-S to avoid excessive resource usage and starvation respectively. As we have mentioned, the ETSI adaptive DCC algorithm is based on LIMERIC~\cite{bansal2013}, and the key difference is the chosen values for their parameters. Another difference is that, although \cite{bansal2013} studies the use of a $G^{+}_{max}$ = $G^{-}_{min}$ to limit the maximum variation of the $\delta_{offset}$, this is not used in much of the LIMERIC simulation work.   

The original LIMERIC paper~\cite{bansal2013} analyses in depth the convergence of the algorithm. Based on this analysis, we can state that in the ETSI adaptive DCC algorithm, the theoretical $\delta$ to which each ITS-S converges is $\delta_{conv}$ as given by equations~\ref{eq:conv} and \ref{eq:delta_conv}:

\begin{equation}
\label{eq:conv}
conv=min\left\{\frac{G^{+}_{max}}{\alpha},\frac{\beta \times CBR_{t}}{\alpha + K \times \beta} \right\} 
\end{equation}

where $K$ is the number of ITS-Ss sharing the channel, and:

\begin{equation}
\label{eq:delta_conv}
\delta_{conv} = 
\begin{cases}
conv & \text{if } 
\begin{cases}
(\alpha + K\times\beta) < 2 \text{ and}  \\
conv \leq \delta_{max} \text{ and} \\
conv \geq \delta_{min} 
\end{cases} \\ 
\frac{G^{+}_{max}}{\alpha} & \text{if }
\begin{cases}
\frac{G^{+}_{max}}{\alpha} \leq \frac{\beta \times CBR_{t}}{\alpha + K \times \beta} \text{ and} \\
\frac{G^{+}_{max}}{\alpha} \leq \delta_{max} \text{ and}\\
\frac{G^{+}_{max}}{\alpha} \geq \delta_{min}
\end{cases}\\ 
\delta_{max} & \text{if } conv > \delta_{max} \\ 
\delta_{min} & \text{if } conv < \delta_{min} \\[0.5ex]  
\textit{no guaranteed} \\[-1ex]
\textit{convergence} & otherwise 
\end{cases}
\end{equation}

\subsection{The Parameters of the ETSI Adaptive DCC Algorithm}

\begin{figure}[!t]
\centering
\includegraphics[width=0.75\columnwidth]{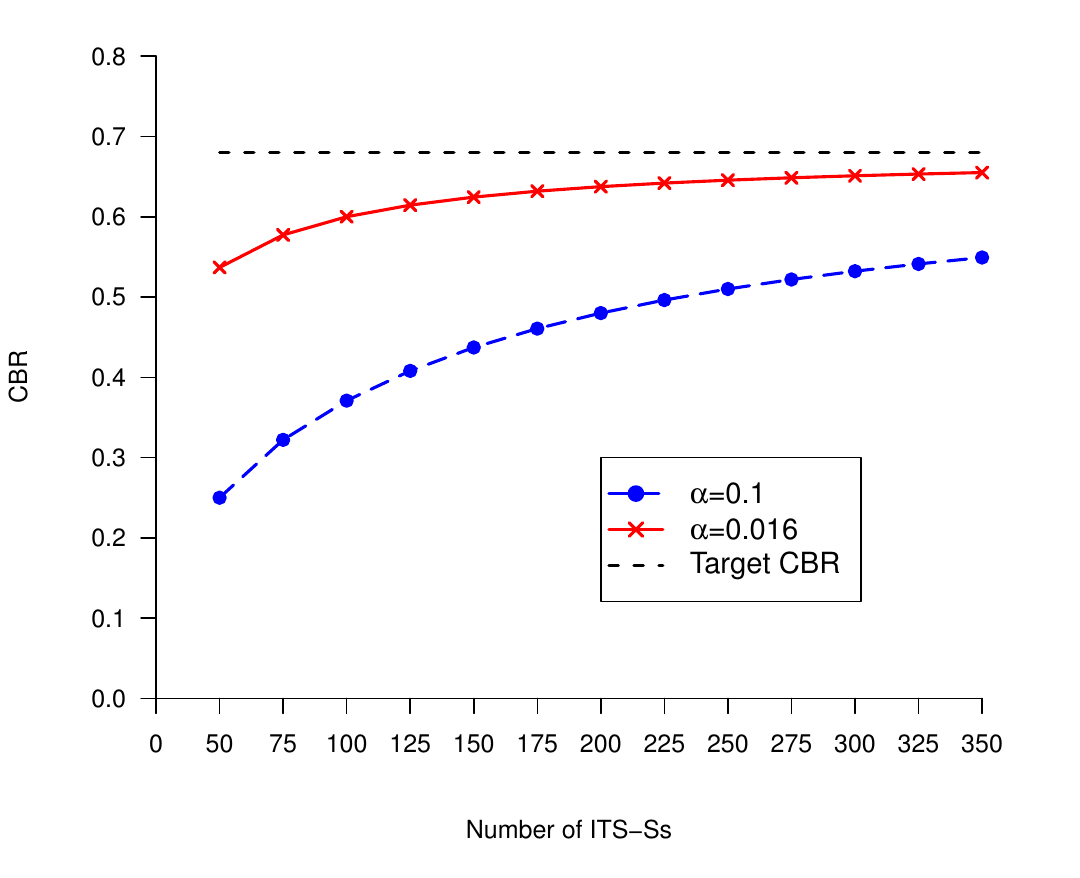}
\caption{Comparison of achieved CBR for $\alpha=0.1$ and $\alpha=0.016$}
\label{fig:alphas}
\end{figure}

The values for the parameters in the ETSI adaptive DCC algorithm (equations \ref{eq:offset}, \ref{eq:delta_wb}, and \ref{eq:delta_n}) are presented in Table~\ref{tbl:parameters_dcc}. It is interesting to understand the rationale for the chosen values for the parameters and, in particular, for $\alpha$ and $\beta$. $\alpha$ has a small value when compared with the value used in LIMERIC papers \cite{bansal2013, rostami2016} in which $\alpha=0.1$. Equation~\ref{eq:conv} means that a smaller $\alpha$ allows to obtain a $CBR$ in convergence closer to $CBR_{t}$. This is shown in Figure~\ref{fig:alphas} where we compare the analytical values of CBR achieved in convergence when different numbers of ITS-Ss share the medium, using the parameters of the ETSI adaptive DCC algorithm but with two different values of $\alpha$: 0.1 and 0.016.

\begin{table}[tb]
\centering
  \caption{Parameter values of the ETSI adaptive DCC algorithm}
  \label{tbl:parameters_dcc}
  \begin{tabular}{|c|c||c|c|}
    \hline
    \textbf{Parameter}  & \textbf{Value} & \textbf{Parameter}  & \textbf{Value} \\
    \hline
     $\alpha$ & 0.016   &   $\delta_{max}$ & 0.03 \\
    $\beta$ & 0.0012  & $\delta_{min}$ & 0.0006 \\
   $CBR_{t}$ & 0.68 &   $G^{+}_{max}$ & 0.0005 \\
  & & $G^{-}_{min}$ & -0.00025 \\  
    \hline
  \end{tabular}
\end{table} 

Regarding the value of $\beta$, in equation~\ref{eq:offset} $\beta$ is used to distribute the difference between the measured $CBR$ and $CBR_{t}$ among all ITS-Ss, i.e., it represents the inverse of the assumed number of ITS-Ss. The problem is that the algorithm has to work for any possible number of ITS-Ss. If $\beta$ is too small compared with the inverse of the real number of \mbox{ITS-Ss} present, the speed of convergence is slower, because each \mbox{ITS-S} thinks that its share of the difference with the $CBR_{t}$ is less than it really is, so more steps of the algorithm are needed to reach convergence. However, large $\beta$ values may result in the algorithm operating in situations in which it does not converge (i.e., permanent oscillatory behavior) and, even if it does, it can create larger oscillations around the convergence point until it reaches it.

The value of $\beta$ chosen by the ETSI DCC algorithm is quite small and equivalent to 833.33 ITS-Ss. In combination with $\alpha=0.016$, it guarantees the convergence of the algorithm because  with $K \geq 1133.33$ ITS-Ss, $\delta$ converges to $\delta_{min}$ and with $K<1133.33$ ITS-Ss, $(\alpha + K \times \beta)<2$. In fact, since the algorithm does not allow $\delta$ to go below $\delta_{min}$, this creates a  limit to the number of ITS-Ss that the algorithm is able to manage (1133.33) without the  $CBR$ starting to go over $CBR_{t}$.  In summary, the  ETSI adaptive DCC algorithm parameters guarantee convergence for any number of ITS-Ss and the achieved $CBR$ in convergence is closer to $CBR_{t}$ than using larger values of $\alpha$.

On the other hand, a small value of $\alpha$ has disadvantages in terms of speed of convergence from high utilization situations. In these situations, collisions hide the excess $CBR$, and $(CBR_{t} - CBR_{s}(n))$ in equation~\ref{eq:offset} is a low estimation of how much we have to decrease $\delta$. Since, in equation~\ref{eq:delta_wb} the previous value of $\delta$ is multiplied by $(1-\alpha)$, this basically means that each \mbox{ITS-S} is relinquishing preventively part of its $\delta$. The relinquished $\delta$ helps to converge from high utilization situations, but the impact of this effect is proportional to the value of $\alpha$.

Another related problem with a small $\alpha$ is fairness. This arises, for example, when two groups of ITS-Ss merge, and the ITS-Ss in each group start with a different $\delta$. In this situation, we would like all ITS-Ss to converge to the same $\delta$ as soon as possible. In equation~\ref{eq:delta_wb}, $\delta_{offset}$ is the same for all ITS-Ss, as it depends on the difference between $CBR_{t}$ and the measured $CBR$, which should be the same for all ITS-Ss sharing the channel. So, to balance the $\delta$ of the ITS-Ss, we depend on the $(1-\alpha) \times \delta(n-1)$ term, which basically makes the \mbox{ITS-Ss} with larger $\delta$ lose more of their $\delta$. As mentioned before, the $\alpha \times \delta(n-1)$ term means that each ITS-S is relinquishing preventively part of its $\delta$, which helps to welcome new \mbox{ITS-Ss} and allow them to achieve a fair slice of the medium. But this effect is diminished if $\alpha$ is small.   

\section{Dual-$\alpha$ DCC: A Modification to the ETSI Adaptive DCC Algorithm}

The ETSI adaptive DCC algorithm performs worse than the original configuration of the LIMERIC algorithm in situations of convergence from high utilization and in transitory situations with a mix of ITS-Ss using different $\delta$. The main reason is the small value of $\alpha$ in the ETSI adaptive DCC algorithm ($\alpha=0.016$) compared with the $\alpha$ used in LIMERIC papers ($\alpha=0.1$). Nevertheless, as we have described, using a small value of $\alpha$ has advantages in terms of the achieved utilization of the medium (closer to $CBR_{t}$).

To solve this trade-off in the value of $\alpha$, we propose to use a dual $\alpha$ value according to equation~\ref{eq:newalpha}:

\begin{equation}
\label{eq:newalpha}
\alpha(n) =
\begin{cases}
\alpha_{high} & \text{if }  
 (\delta(n-1) - \delta_{\alpha_{low}}(n)) > th  \\ 
\alpha_{low} & otherwise
\end{cases}
\end{equation}

where $\alpha(n)$ is the $\alpha$ to use in equation~\ref{eq:delta_wb} for step \textit{n}, $ \delta_{\alpha_{low}}(n)$ is a $\delta$ calculated using equation~\ref{eq:delta_n} with $\alpha=\alpha_{low}$, and $th$ is a threshold whose function is explained next. The parameters for this Dual-$\alpha$ algorithm (equations \ref{eq:cbr}, \ref{eq:offset}, \ref{eq:delta_wb}, \ref{eq:delta_n}, and \ref{eq:newalpha}) are presented in Table~\ref{tbl:parameters_modified_dcc} (parameters not included are equal to those in Table~\ref{tbl:parameters_dcc}). The key idea is to use one value of $\alpha$ when $\delta$ is decreasing, and another when it is increasing. $th$ is a threshold intended to improve stability when the algorithm is close to convergence. Its value is heuristic and it has been shown to improve stability in the whole range of \mbox{ITS-S} densities covered by the standard. A smaller threshold can result in a potential fluctuation of less than 5\% around the convergence value of $\delta$, which is not necessarily a problem because $\delta$ and $CBR$ are, in any case, dynamic. 

\begin{table}[tb]
\centering
  \caption{Parameter values of the Dual-$\alpha$ DCC algorithm}
\label{tbl:parameters_modified_dcc}
  \begin{tabular}{|c|c|}
    \hline
    \textbf{Parameter}  & \textbf{Value} \\
    \hline
     $\alpha_{low}$ & 0.016 \\
    $\alpha_{high}$ & 0.1 \\   
   $th$ & 0.00001 \\
    \hline
  \end{tabular}
\end{table} 

\section{Evaluation of Scenarios of Slow Convergence and Unfairness}

This section shows, using numerical analysis and simulations, the improvement in speed of convergence and fairness achieved by our proposed Dual-$\alpha$ DCC algorithm. 

\subsection{Scenarios}

We analyze two scenarios. In the convergence scenario, we have  a number of ITS-Ss that start with a value of $\delta$ equal to $\delta_{max}$. This could happen, for example, in a traffic jam, when vehicles are still and reduce their message rate, so they perceive a free medium which increases the potential $\delta$ they can use. When movement resumes and message rate increases, cars will use the calculated $\delta$, but since there can be many vehicles, the DCC algorithm needs to converge to a smaller value of $\delta$.  

In the second scenario, we have two groups of ITS-Ss sharing the medium. At the beginning, the ITS-Ss in each group have a different $\delta$. This could happen, for example, when two groups of vehicles merge in a junction. The $\delta$ for all ITS-Ss must converge to the same value.

\subsection{Numerical Results}

Table~\ref{tbl:convergence} presents the time the ETSI adaptive DCC algorithm and our Dual-$\alpha$ DCC algorithm need to reduce the occupancy in the channel below the 0.68 target value (i.e., time to achieve the first $CBR$ value below 0.68) when the ITS-Ss in a group (ranging from 100 to 1100 ITS-Ss) start sending at the rate allowed by $\delta_{max}$. As we can see in Table~\ref{tbl:convergence}, the time needed by the Dual-$\alpha$ algorithm is between  25\% and 35\% of the time needed by the ETSI adaptive DCC algorithm.   

 \begin{table}[tb]
\centering
  \caption{Speed of convergence}
\label{tbl:convergence}
  \begin{tabular}{|c|c|c|}
    \hline
    \textbf{Number of ITS-S}  & \textbf{ETSI DCC $<0.68$} & \textbf{Modified DCC $< 0.68$}\\
    \hline
     100  & 9.4s &  2.4s  \\
     300 &    11.8s        &  3.8s   \\
     500 &     12.4s       &    4.2s \\
     700 &     12.6s       &   4.4s  \\
     900 &       12.8s     &    4.4s \\
     1100 &      13s      &     4.6s\\
    \hline
  \end{tabular}
\end{table} 

Table~\ref{tbl:fairness} presents the results for the ETSI adaptive and Dual-$\alpha$ DCC algorithms when two groups of \mbox{ITS-Ss} merge. One group has 25 ITS-Ss starting with a $\delta$=0.0177, which is the convergence value for this group size. We analyze several cases for the size of the second group (from 100 to 1100 \mbox{ITS-Ss}) and, in each one, the ITS-Ss start with a value of $\delta$ equal to the convergence value for a group of the respective size. Therefore, we are representing the situation of two groups of \mbox{ITS-Ss} that have reached convergence separately and then have merged. The performance metrics in the table are: \mbox{\textit{JI-10s}} (the Jain fairness index 10s after the groups merged, computed as $(\sum_{i=1}^{K} \delta_{i})^2/(K \times \sum_{i=1}^{K} \delta_{i}^2)$ with \textit{K} being the total number of \mbox{ITS-Ss}); \textit{$t_{conv}$} (the time needed for the larger group of \mbox{ITS-Ss} to achieve a $\delta$ within $\pm$10\% of the convergence $\delta$ of the merged group); and $<68$ (the time needed to achieve the first $CBR$ value below 0.68). 

The Jain index shows the fairness in the sharing of the medium among the ITS-Ss of the merged group. Lower values represent greater unfairness, and a value of $1.0$ means that the sharing is equal among all \mbox{ITS-Ss}. As an example, with 100 \mbox{ITS-Ss} in the second group, the Jain index, 10s after the two groups merged, is equal to \textit{0.86} with the ETSI adaptive DCC algorithm. This value results from having the \mbox{ITS-Ss} coming from the larger group allowed to send messages at a rate that is 42\% of the sending rate of the ITS-Ss coming from the other group. 
In the Dual-$\alpha$ DCC algorithm, for the same situation, the sending rate in the \mbox{ITS-Ss} coming from the larger group is 91\% of the sending rate of the \mbox{ITS-Ss} coming from the smaller group. The other two metrics can be interpreted as follows: the ITS-Ss coming from the smaller group have larger $\delta$ and are the slowest to converge, but this is not really a problem if it is not harming the other group of \mbox{ITS-Ss} nor congesting the channel, which is the case when \textit{$t_{conv}$} is small and  $<68$ is small. As it can be seen, the performance of the Dual-$\alpha$ DCC algorithm is a great improvement compared with the performance of the ETSI adaptive DCC algorithm. 

 \begin{table}[tb]
\centering
  \caption{Fairness when combining a group of 25 ITS-Ss with a second group of ITS-s}
\label{tbl:fairness}
  \begin{tabular}{|c||c|c|c||c|c|c|}
    \hline
    \multirow{2}{0.9cm}{\textbf{ITS-Ss}}  & \multicolumn{3}{|c||}{\textbf{ETSI DCC}} & \multicolumn{3}{|c|}{\textbf{Dual-$\alpha$ DCC}} \\
    & \textbf{JI-10s} & \textbf{$t_{conv}$} & $<68$ & \textbf{JI-10s} & \textbf{$t_{conv}$} & $<68$ \\
    \hline
     100  & 0.86  & 19.4s  & 2s & 0.998 & 6s  & 0.6s \\
300  &  0.53  &  22.2s&  1s & 0.994 & 3.8s & 0.6s \\
500  &  0.39  & 22.4s &  1.2s & 0.988 & 3.4s & 0.4s\\
700  &  0.34 &  20.6s & 4.6s  &  0.980 & 3.4s  & 1s \\
900  &  0.39 &  16s &  8.4s & 0.974 & 3s & 2s \\
1100  &  0.70 & 0s & 17.8s &  1 & 0s & 4.8s \\
    \hline
  \end{tabular}
\end{table}

\subsection{Simulation Results}
\begin{figure}[!t]
\centering
\includegraphics[width=0.70\columnwidth]{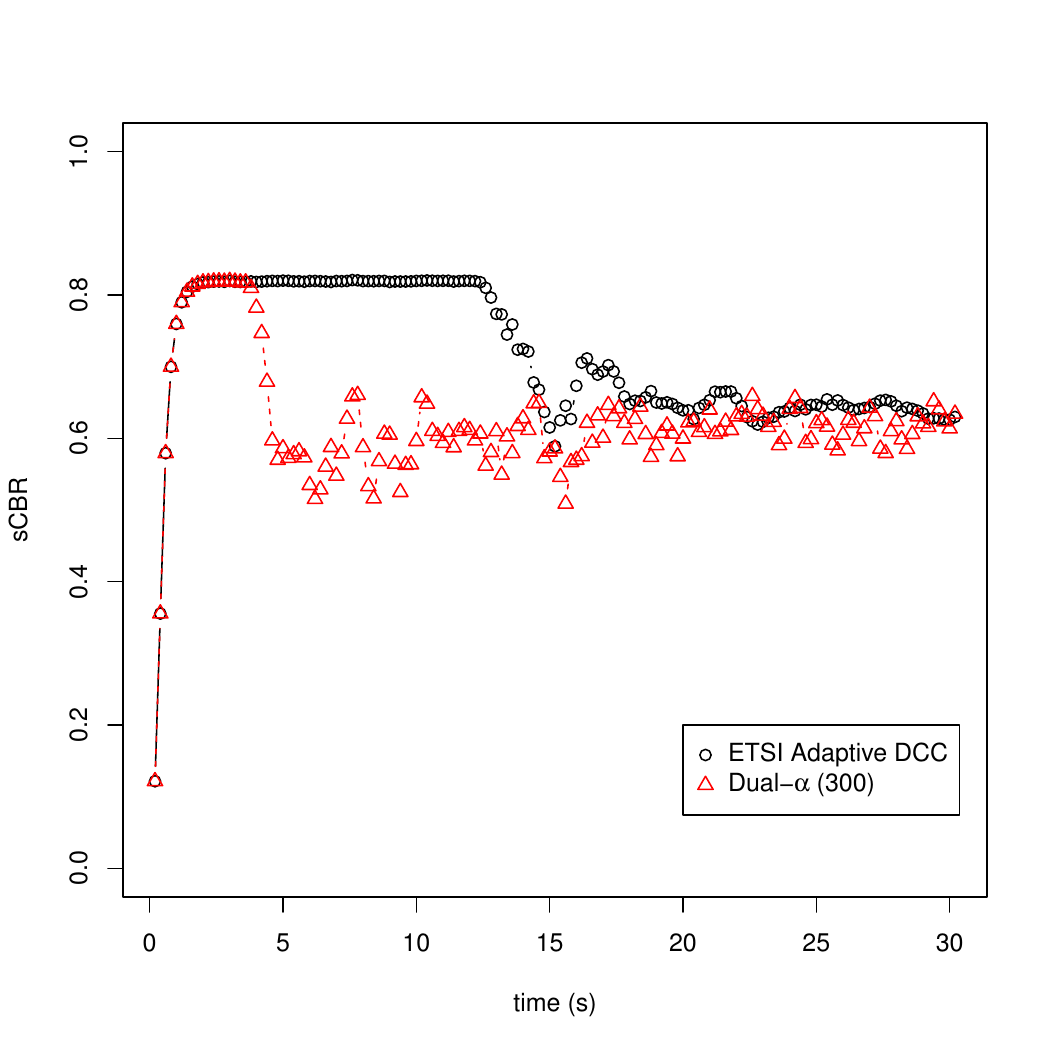}
\caption{Evolution of $CBR_{s}$ for 300 ITS-Ss starting in $\delta=0.03$.}
\label{fig:convergence}
\end{figure}

\begin{figure}[!t]
\centering
\includegraphics[width=0.70\columnwidth]{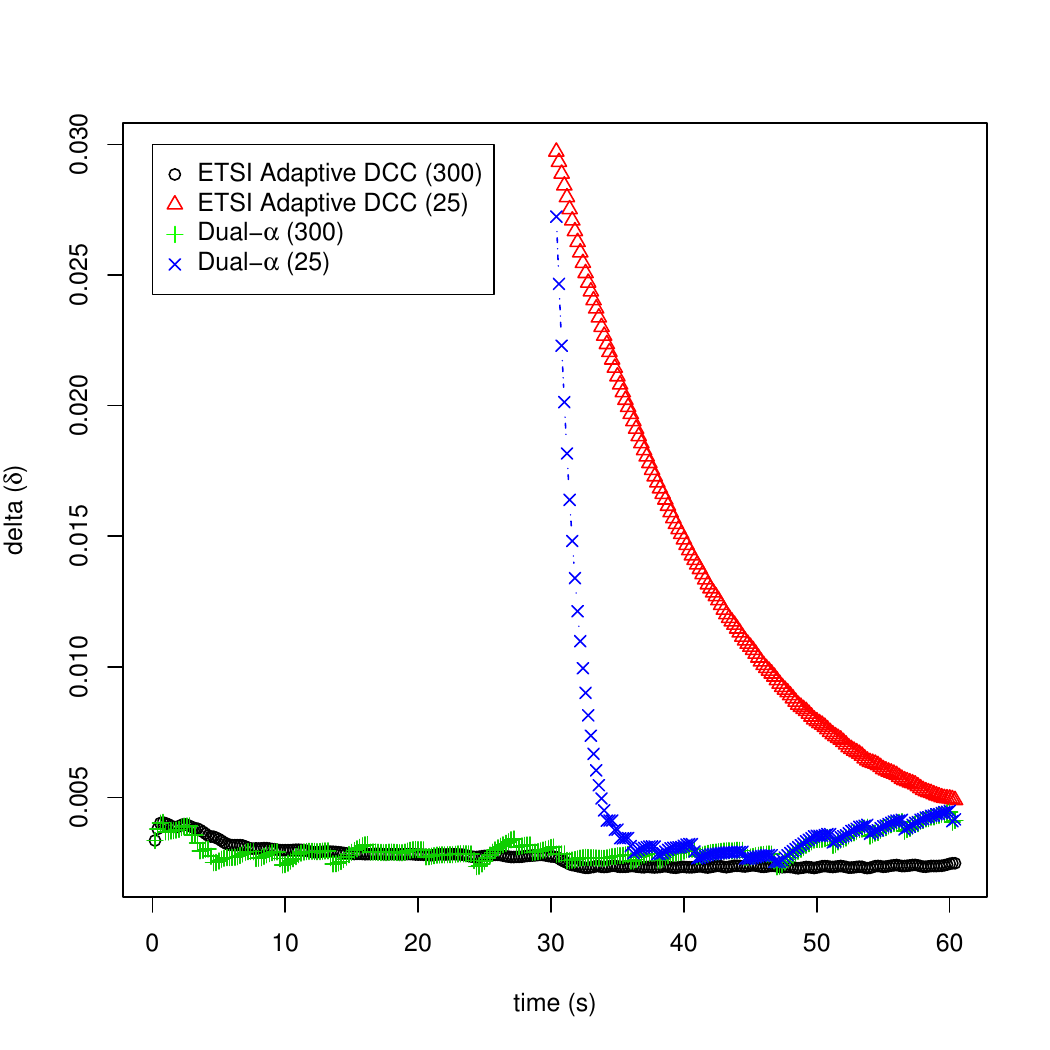}
\caption{Evolution of $\delta$ for vehicles in two different groups that merge.}
\label{fig:fairness}
\end{figure}

We have obtained some simulation results to confirm the numerical analysis. Our future work includes more extensive simulations to evaluate the proposal. For the simulations we have used Artery~\cite{riebl2015}, 
a simulation framework for ETSI \mbox{ITS-G5}, and using the Two-Ray Interference model of Veins~\cite{sommer2011}. In the first scenario, 300 static ITS-Ss start with a $\delta$ equal to $\delta_{max}$ (0.03) and Figure~\ref{fig:convergence} shows the evolution of $CBR_{s}$ for the ETSI adaptive DCC algorithm and for the Dual-$\alpha$ DCC algorithm. In the second scenario, we start with 300 vehicles and, after 30s, 25 vehicles join the scenario with a starting $\delta$ equal to the convergence value of the $\delta$ in a group of 25 vehicles. Figure~\ref{fig:fairness} shows the evolution of $\delta$ for two vehicles, one in each group, for the two versions of the DCC algorithm. The simulations results, therefore, confirm our numerical analysis of the advantages of the Dual-$\alpha$ algorithm in terms of convergence and fairness in transitory situations.

\ifCLASSOPTIONcaptionsoff
  \newpage
\fi

\bibliographystyle{IEEEtran}
\bibliography{dcc_abrv}

\end{document}